\documentclass[twocolumn,showpacs,preprintnumbers,amsmath,amssymb,floatfix]{revtex4}

\usepackage{epsfig,psfrag}
\usepackage{dcolumn}
\usepackage{bm}
\usepackage{graphicx}
\usepackage{color}

\newcommand{\ket}[1]{|#1\rangle}
\newcommand{\bra}[1]{\langle#1|}
\newcommand{\mel}[3]{\langle#1|#2|#3\rangle}

\newcommand{\be}{\begin{equation}} 
\newcommand{\ee}{\end{equation}}
\newcommand{\bea}{\begin{eqnarray}} 
\newcommand{\eea}{\end{eqnarray}}

\begin{document}

\title{Current-induced non-adiabatic spin torques
 and domain wall motion with spin relaxation in a ferromagnetic
metallic wire}
\author{M. Thorwart and R. Egger}
\affiliation{Institut f\"ur Theoretische Physik,
Heinrich-Heine-Universit\"at D\"usseldorf, D-40225  D\"usseldorf, Germany}

\date{\today}

\begin{abstract}
Within the $s$-$d$ model description, we derive the 
current-driven spin torque in a ferromagnet, taking explicitly into account
a spin-relaxing Caldeira-Leggett bath coupling to the $s$-electrons.
We derive Bloch-Redfield equations of motion for the 
$s$-electron spin dynamics, and formulate a gradient expansion scheme to
obtain non-adiabatic (higher-order) corrections to the well-known
adiabatic (first-order) spin torque.
We provide simple analytical expressions for the second-order spin torque.
The theory is applied to current-driven domain wall motion. 
Second-order contributions imply a deformation of a 
transverse tail-to-tail domain wall.  The wall center still moves with a
constant velocity that now depends on the spin-polarized current in 
a non-trivial manner.
\end{abstract}

\maketitle

\section{Introduction}

Many recent experiments have brought to attention the intricate
and rich interplay between spin-dependent transport and 
the magnetization dynamics in ferromagnetic (FM) 
materials \cite{fabian,review,brataas}.
A particularly interesting issue
concerns the {\sl spin torque\/} ${\bf T}$
 \cite{berger,bjz,lzad,macdon,tatara1,piechon} 
exerted by conduction electrons on the FM magnetization, $-M_s {\bf n}$,
where ${\bf n}(x,t)$ is a space-time dependent unit vector and 
$M_s$ is assumed constant. 
The spin torque is crucial for the understanding
of many topics of present scientific and technological interest, including
current-driven domain wall (DW) motion \cite{dw},
magnetization precession and switching in multilayer geometries \cite{ralph},
and spin transport in general \cite{spintransport}.
Here, we study the spin torque exerted by a steady-state
dc spin current $J_s$ [for the definition of $J_s$, see Eq.~(\ref{jdef})] 
in the presence of spin relaxation.  Common 
microscopic mechanisms responsible for spin relaxation are, for instance,
provided by magnetic impurities or the various sources for 
spin-orbit scattering, 
but we shall focus on generic features and study a phenomenological
Caldeira-Leggett-type heat bath describing random magnetic fields
with Gaussian statistics which are responsible for spin relaxation.  

As has been discussed recently, spin relaxation is essential
in determining the non-adiabatic corrections to 
the well-known {\sl adiabatic spin torque} \cite{bjz,lzad} 
\begin{equation} \label{ad1}
{\bf T}_{ad} = - J_s \partial_x {\bf n} .
\end{equation}
It is common  practice to call
all terms beyond Eq.~(\ref{ad1}) `non-adiabatic' (but see below).
In particular, Zhang and Li \cite{li} predicted a leading non-adiabatic
spin torque contribution 
\begin{equation}\label{liz}
{\bf T}_{ZL} =  \beta J_s {\bf n}\times \partial_x {\bf n} ,
\end{equation}
which was confirmed by later work
\cite{barnes,gerrit,tatara2,tatara3,duine}, 
although some questions were raised in Ref.~\cite{stiles}.
Here the dimensionless $\beta$ parameter is given as
$\beta=\hbar/\Delta\tau_{s}$, where $\Delta$ is the FM exchange
splitting and $\tau_{s}$ denotes a spin relaxation time.
 Although $\beta$ is generally small, ${\bf T}_{ZL}$
can profoundly influence many of the above-mentioned phenomena. 
Recently, a complicated and highly non-local form of the spin transfer torque
including non-adiabatic corrections due to fast varying magnetic
textures has been proposed \cite{tatara4}. This form includes
indirect Ruderman-Kittel
 exchange interaction processes \cite{rkky,simanek} between the
localized spins mediated by itinerant electrons. A scattering approach
has been employed in Ref.\  \cite{dugaev} 
to calculate the  spin torque in the limit of  a sharp DW. However, it
is difficult to deal with spin relaxation in that formalism. 
Strong enhancement of the spin torque and of the domain wall mobility
due to spin-orbit coupling in magnetic II-V semiconductors has been
reported in Ref.\  \cite{brataas2}. 

In this work, we provide a general derivation of the spin torque 
within the $s$-$d$ model \cite{early,yosida}, see Sec.~\ref{sec2}.  
The self-consistent Stoner model gives very similar results for 
this problem but is technically somewhat more demanding 
\cite{gerrit,tatara2,duine}.
While spin torques in the presence of spin relaxation have so far been 
studied mostly for small perturbations around a homogeneously
 magnetized ferromagnet \cite{li,gerrit,tatara2}, 
we here describe the theory for the leading nonadiabatic corrections for
non-homogenous (both in space and time) magnetization profiles.
For these corrections and not extremely large applied currents,
the $s$-$d$ model is expected to provide a reasonable description.
We note in passing that recently a 
simple classical model of non-adiabatic
current-induced spin torques has been proposed \cite{viret}.
In our work, spin relaxation is included within the weak-coupling 
Bloch-Redfield approach \cite{louisell,abragam,slichter,may}, 
where the $s$-electrons are coupled to a 
spin-relaxing environment modelled by a phenomenological harmonic oscillator 
bath \cite{weiss}, see Sec.~\ref{rel}. 
Moreover, the itinerant electron spins are influenced by a space-time
dependent magnetic field produced by the localized electron spins. 
This external field also determines the stationary
state to which the itinerant spins relax.
Validity of the Bloch-Redfield approach requires  weak coupling 
of the $s$-electrons  to the bath, i.e., sufficiently long spin relaxation
times $\tau_s$, and  a bath memory time $\tau_c$ short
compared to all other relevant timescales. 
Both conditions hold in most cases of practical interest.

Based on this formalism, we then determine the spin torque ${\bf T}$
acting on the FM magnetization ${\bf n}(x,t)$
within a gradient expansion around the adiabatic limit, see
Sec.~\ref{sec23}.  We illustrate the method by computing
all first- and second-order derivative contributions to the spin torque,
i.e., all terms that involve $\partial_{x,t}{\bf n}, \partial_{x,t}^2{\bf n},
\partial_x\partial_t {\bf n}$, or products of first-order derivatives.
The first-order terms provide the leading terms in this expansion,
which we call the {\sl adiabatic torque}.  Note that our convention
disagrees with common usage but emphasizes that these terms completely
dominate the spin torque for slowly varying ${\bf n}(x,t)$.
Indeed, a first-order expansion recovers both Eq.~(\ref{ad1}) and (\ref{liz}).
This shows that the `non-adiabatic' term (\ref{liz})
is really a consequence of spin relaxation rather than being related
to higher-order terms in the gradient expansion \cite{li,gerrit,tatara2}.  
Our approach then allows to systematically  
evaluate higher orders in an expansion around the adiabatic limit.
It should be stressed, however, that the $s$-$d$ model may become
difficult to justify for extremely sharp or fast features in ${\bf n}(x,t)$
(e.g. very narrow domain walls), and in practice this expansion is only
useful for the first few orders.

Here we will focus in detail on the 
second-order spin torque contributions which turn out to be of
relatively simple form.  
To illustrate some physical consequences of these new 
terms, in Sec.~\ref{sec3} we address the problem of 
DW motion driven by a spin-polarized current
 for the case of a tail-to-tail transverse DW.
We find that the second-order contribution leads to a deformation of
the adiabatic DW shape. Nevertheless, under steady state conditions
the DW center moves uniformly with constant terminal velocity $V_{DW}$. 
Importantly, this velocity depends non-trivially on the spin
current density $J_s$. After following the known linear behavior of the
velocity found from the first-order calculation for small $J_s$,
$V_{DW}$ does not grow further with $J_s$ but instead  starts to
decrease.  We expect that when including all orders,
this would indicate saturation behavior. Our results seem in qualitative
agreement with experimental observations, where measured 
DW velocities are typically smaller than those predicted by
first-order calculations \cite{meier}. Moreover, the deformation of the 
DW profile is also observed in
 experiments and can even induce a transformation
to a completely different DW type \cite{klaui}. 
While the quantitative description of experiments on
current-driven DW motion is likely to require more
refined models, our principal aim is to provide  general
expressions for the leading corrections to the known spin torque
expressions (\ref{ad1}) and (\ref{liz}), and to illustrate 
typical effects caused by these new terms.
The practical importance of such effects was also revealed by recent 
numerical studies \cite{waintal,thiaville,kramer}.

Our paper then closes by offering
some concluding remarks in Sec.~\ref{sec4}.  The derivation
of the relaxation kernel under the Bloch-Redfield approach
can be found in the Appendix.
Throughout the paper, we put $\hbar=k_B=1$.

\section{Model}
\label{sec2}

We consider an (infinitely long)
FM wire with homogeneous magnetization
along the cross-section, ${\bf n}={\bf n}(x,t)$, where
$x$ is the longitudinal direction.  This simplification
allows us to work with an effectively one-dimensional (1D) 
theory. For thin FM nanowires with only a few occupied transverse sub-bands,
this description is directly relevant \cite{balents}, but for
wider wires, nontrivial transversal magnetization profiles  (such 
as vortex walls) are thereby excluded. 
Moreover, the 1D model also neglects spin waves 
propagating in the transverse direction that can be excited in wider wires.
We note in passing that in strictly 1D metallic wires, 
electron-electron interactions
can cause non-Fermi liquid effects \cite{gogolin} that could also influence
domain-wall motion \cite{araujo}.
However, such effects are not studied in what follows.

Adopting the widely used $s$-$d$ model for itinerant FMs 
\cite{review,brataas,early,yosida}, the relevant dynamical
degrees of freedom are (i) the FM magnetization ${\bf n}(x,t)$
associated with localized $d$-electrons,  and (ii) the 
conduction electron spin current density ${\bf J}(x,t)$ and 
spin density
${\bf s}(x,t)$, respectively, describing the
delocalized $s$-electrons.
The FM magnetization then obeys the Landau-Lifshitz-Gilbert (LLG)
equation \cite{review}
\begin{equation}\label{llg}
\partial_t {\bf n} = -\gamma_0 
{\bf n}\times {\bf B}_{\rm eff}[{\bf n}] + \alpha {\bf n}
\times \partial_t {\bf n} + {\bf T}, 
\end{equation}
where  $\gamma_0 {\bf B}_{\rm eff}$ 
 includes external, anisotropy, and exchange magnetic
fields unrelated to the coupling to $s$-electrons.
The Gilbert damping parameter $\alpha$ provides
a phenomenological description of dissipative influences on ${\bf n}$
(again in the absence of $s$-electrons).
Our principal aim is to find the spin torque ${\bf T}$ 
when the FM carries a (spin-polarized) current and both spin relaxation 
and magnetic texture (spatio-temporal variations
of ${\bf n}$) are present.

Let us then address the conduction electron degrees of freedom.
Expressed in terms of right- and left-moving
($p=R/L$) quasi-particle annihilation operators 
$c_{p,\sigma}(x)$ for 
spin $\sigma=\uparrow,\downarrow$, the 1D spin and spin current density are
defined as
\begin{equation} \label{spin}
{\bf s}(x)={\bf J}_R(x)+{\bf J}_L(x),\quad
{\bf J}(x)=v({\bf J}_R(x)-{\bf J}_L(x) ),
\end{equation}
where $v$ is the spin velocity and the chiral currents are
\begin{equation}\label{spincurr}
{\bf J}_{R/L}(x) = \frac12 : c^\dagger_{R/L}(x)
\vec\sigma c^{}_{R/L}(x):.
\end{equation}
Here colons denote normal ordering, spin indices are left implicit,
and $\vec\sigma$  are standard Pauli matrices acting in spin space.  
The currents (\ref{spincurr}) obey the 
$SU(2)$ Kac-Moody algebra \cite{gogolin}.
Within this 1D description, the low-energy Hamiltonian
describing the conduction electrons is universal and given by
\begin{equation}\label{h1d1}
H_0  =  -iv\sum_{p=R/L=\pm}  p \int dx \ c_p^\dagger \partial_x c_p^{}.
\end{equation}
The spin sector of $H_0$ can now be expressed
directly in terms of the currents (\ref{spincurr}), 
and completely decouples from the charge sector.
(In a strictly 1D wire, this procedure works even when forward-scattering
electron-electron interactions are included \cite{gogolin}.)
The resulting Sugawara representation of the spin part of $H_0$ (the 
charge part is irrelevant and omitted here) is 
\begin{equation} \label{h1d}
H_{0}  = \frac{v}{4}\int dx : {\bf s} \cdot {\bf
s} + \frac{1}{v^2} {\bf J} \cdot {\bf J} :.
\end{equation}
Within the $s$-$d$ model, $s$-electrons are coupled
to the time-dependent magnetization ${\bf n}(x,t)$ only by exchange processes.
Effectively, the magnetization then acts  on the $s$-electrons
as a time-dependent external field,
\begin{equation}\label{hex}
H_{ex}(t) =  \Delta \int dx \ {\bf s}(x) \cdot {\bf n}(x,t),
\end{equation}
where $\Delta$ is proportional to the 
FM exchange coupling and can have either sign. 
For the resulting `system' (the $s$-electrons), 
we then arrive at the Hamiltonian $H_S(t)=H_0+ H_{ex}(t)$. 
Equation (\ref{hex}) now generates a spin torque entering Eq.~(\ref{llg}),
\begin{equation}\label{torq}
{\bf T} = {\bf n} \times \left(-\frac{\delta H_{ex}}{\delta {\bf
n}}\right) =  -\Delta {\bf n} \times {\bf s} .
\end{equation}
Our task is then to compute the spin density ${\bf s}(x,t)$ 
in the presence of an external dc spin current $J_s$ and 
for a given dynamical magnetization profile ${\bf n}(x,t)$.
We mention in passing that we do {\sl not\/} perform a unitary
transformation into an adiabatic reference frame where the
local quantization axis for the itinerant electrons is
aligned with ${\bf n}(x,t)$.  Albeit this transformation has
been used in many papers on the subject, see e.g. Ref.~\cite{shibata} 
and references therein,
we believe that it does not offer advantages in the 
presence of spin relaxation.

In order to enforce a finite spin current, in principle (i)
we need to resort to a non-equilibrium formalism. In addition,
(ii) we need a proper description of the relaxation processes
driving the momentary state to a stationary state. Let us first
discuss issue (i), where a simple boundary condition can be used
instead of the full dynamical formalism, see also Ref.~\cite{balents}.
We want to describe a steady-state situation with
externally imposed constant spin current 
\begin{equation} \label{jdef}
J_s =  \frac{PI}{eA},
\end{equation}
expressed in terms of the charge current $I$ flowing 
through a wire of cross-section $A$ with spin polarization $0<P\leq 1$.
Since this stationary spin current is externally enforced, we
include it as a boundary condition in the equation
of motion of the itinerant electrons derived below. This boundary 
condition has to be imposed far away from any magnetic texture (i.e., at
$x\to \mp\infty$), where it 
implies a constant spin current ${\bf J}=J_s {\bf n}$. 
Only around, say, a DW center, the true spin current density deviates from 
${\bf J}(x,t)=J_s{\bf n}(x,t)$.  
Coming to issue (ii), it is then tempting to identify
the steady state to which an actual spin current configuration
tends to relax with ${\bf J}(x,t)=J_s{\bf n}(x,t)$. 
This indeed we find, see Sec.~\ref{rel}.
The Heisenberg equations of motion,
$\partial_t {\bf J}_{R/L}= i[H_S(t),{\bf J}_{R/L}]$, 
now yield operator equations for the spin density and the spin current,
\begin{eqnarray}\label{eqmsja}
 \partial_t {\bf s} + \partial_x {\bf J} & = &  
 - \Delta {\bf s} \times  {\bf n}+  
{\bf \Gamma}_s , \\ \label{eqmsjb}
 \partial_t {\bf J} + v^2 \partial_x {\bf s} & = &  
 - \Delta {\bf J} \times {\bf n} + {\bf \Gamma}_J  .
\end{eqnarray}
The relaxation terms ${\bf \Gamma}_{s/J}[{\bf s},{\bf J}]$ 
due to the coupling of the $s$-electrons to the spin-relaxing
environment are specified in Sec.~\ref{rel}.
Far away from magnetic textures,
both the derivative terms and the relaxation terms 
${\bf \Gamma}_{s,J}$ are irrelevant, and  
taking ${\bf s}=s_0 {\bf n}$ and ${\bf J}= J_0{\bf n}$ 
with arbitrary constant coefficients $s_0$ and $J_0$
solves these equations. 
The boundary condition on ${\bf J}$ discussed above then
enforces $J_0=J_s$. 
To determine $s_0$, one has to find the stationary state of
the equation of motion (\ref{eqmsja}) in the presence of spin relaxation.
We here anticipate the result $s_0= -\chi_s \Delta$ derived in 
 Sec.~\ref{rel}, see Eq.~\eqref{rels}, where
 $\chi_s=(2 \pi v)^{-1}$ is the spin susceptibility. 

To summarize our discussion,
we seek solutions to Eqs.~(\ref{eqmsja}) and (\ref{eqmsjb}) 
of the form
\begin{eqnarray} \label{ansatz}
{\bf J}(x,t) & = & J_s {\bf n}(x,t) + \sum_{k=1}^\infty {\bf J}_k (x,t) 
 , \\  \label{ansatz2}
{\bf s}(x,t) & = & -(\chi_s \Delta ) {\bf n}(x,t) + \sum_{k=1}^\infty {\bf s}_k(x,t) ,
\end{eqnarray}
where the $k$th-order terms contain $k$th-order spatio-temporal
derivatives of ${\bf n}$, or respective products of 
lower-order derivatives.  Such an expansion provides a 
systematic way to classify deviations from the adiabatic limit
where only the $k=1$ terms are retained.
However, as we have discussed above, in practice
only the first few orders in this expansion can be 
reliably extracted under the $s$-$d$ model description.

\section{Spin relaxation} \label{rel}

For a description of the spin-relaxing environment
acting on the spin dynamics of the $s$-electrons,
we employ a Caldeira-Leggett system-bath approach \cite{weiss},
where the $s$-electron spin density and spin current
are linearly coupled to fluctuating magnetic fields 
${\bf B}_{s,J}(x)$,
\begin{equation} \label{hsb}
H_{SB}  = \int dx  \ [{\bf B}_s(x,t) \cdot {\bf s}(x) + 
{\bf B}_J(x,t) \cdot {\bf J}(x)].
\end{equation}
These fields provide a simple phenomenological modelling of the
effects of random magnetic impurities and of spin-orbit scattering.
The {\sl Ansatz} (\ref{hsb}) assumes `soft' forward scattering due to
the fluctuating fields, where
right- and left-moving fermions retain their chirality.  
Backward scattering processes, for instance associated with 
elastic potential scattering \cite{balents}, will modify our
quantitative conclusions, as discussed in more detail below.
Note also that the fluctuating
fields are assumed to solely couple to $s$-electrons, 
and the FM magnetization 
${\bf n}(x,t)$ is only affected indirectly via its exchange coupling to
the $s$-electron spin density ${\bf s}(x,t)$.  
The Hamiltonian $H_B$ describing the uncoupled fields 
${\bf B}_{s,J}$ corresponds to a suitable bath of harmonic
oscillators \cite{weiss}.  With $\overline{\langle {\bf B}_{s,J} \rangle}
=0$, where $\overline{\langle O\rangle}$ denotes the equilibrium quantum-statistical
average of some operator $O$ with respect to $H_B$, 
all relevant properties are thus encoded by specifying their
two-point correlation functions. 
The total Hamiltonian is then $H_{tot}(t)=H_{S}(t) +
H_{SB}+H_B$, with $H_{S}(t)=H_0+H_{ex}(t)$ as
 given in Eqs.~(\ref{h1d}) and (\ref{hex}). 

To describe the effects of Eq.~(\ref{hsb}) on the $s$-electron
spin dynamics, we adapt the textbook weak-coupling
Bloch-Redfield approach \cite{louisell,abragam,slichter,may}
to a field theory as required here.  For the convenience of the
interested reader, we provide a detailed discussion of the 
Bloch-Redfield equations in the Appendix.
The basic assumptions underlying this approach are (i) 
short bath correlation times, i.e., the 
two-point correlation functions of ${\bf B}_{s,J}(x,t)$
vanish on a timescale $\tau_c$ small compared to all
relevant other timescales,  and (ii) 
weak system-bath coupling $H_{SB}$.
Note that $H_{SB}$ in Eq.~(\ref{hsb}) is of the general 
form considered in the Appendix,
see Eq.~(\ref{hsbapp}).  This implicitly also requires
that we stay not too far away from the adiabatic limit.
Following the steps in the Appendix,
we obtain the relaxation term entering Eq.\  (\ref{eqmsja}) in the form
\begin{equation}\label{rels}
{\bf \Gamma}_s = - \frac{{\bf s}(x,t)+(\chi_s \Delta) {\bf n}(x,t)}{\tau_s} \, ,
\end{equation}
where $\chi_s=(2 \pi v)^{-1}$ and
$\tau_s$ is the spin relaxation time.  Here $\tau_s$ is
assumed uniform in the $x,y,z$ directions; 
the generalization is discussed in the Appendix.
Since ${\bf n}(x,t)$ acts effectively as a space-time dependent magnetic
field on the spin density ${\bf s}$, see Eq.\ (\ref{hex}), it directly appears
in the stationary state approached in the relaxation term (\ref{rels}).
In a similar manner, we obtain 
\begin{equation} \label{relj}
{\bf \Gamma}_J = - \frac{{\bf J}(x,t)-J_s {\bf n}(x,t)}{\tau_J} \, .
\end{equation} 
For the simple system-bath
coupling (\ref{hsb}), which disregards backscattering processes, 
the relaxation times $\tau_s$ and $\tau_J$ turn out to be equal. 
In the following, we shall allow for different relaxation times
$\tau_J\neq \tau_s$,  which results when allowing for
more general models involving backscattering processes \cite{balents},
i.e., when elastic disorder is present. 

Before discussing the resulting spin torque, let us briefly comment on
the validity regime of our description.  Both the derivation of the
relaxation terms, see Eqs.~(\ref{rels}) and (\ref{relj}), and
the iterative procedure for solving the resulting equations, 
see Sec.~\ref{sec23}, suppose that we are not too far away from
the adiabatic limit.  By that we mean that the space (time) variation
of the magnetization direction
${\bf n}(x,t)$ is sufficiently smooth (slow) when
compared to a characteristic lengthscale $\lambda$ (time scale $\lambda/v$).
This `spin transport lengthscale' has been estimated by Zhang and Li, 
see Ref.~\cite{li}.  Quantitatively, we suppose that both dimensionless
non-negative functions
$\lambda |\partial_x {\bf n}(x,t)| \ll 1$ and 
$(\lambda/v)|\partial_t {\bf n}(x,t)|\ll 1$.  Under these
conditions, an iterative expression like Eq.~(\ref{tanz}) 
implies a well-behaved perturbation series, where higher and higher orders
give smaller and smaller corrections.

\section{Spin torque}
\label{sec23}

We next discuss how to extract the spin torque from
an iterative solution of Eqs.~(\ref{eqmsja}) and (\ref{eqmsjb}).
To that end, we perform a gradient expansion 
indicated by the series in Eqs.~(\ref{ansatz}) and (\ref{ansatz2}).
This gradient expansion then generates a series for the spin torque,
\begin{equation}\label{tanz}
{\bf T}(x,t) = \sum_{k=1}^\infty {\bf T}_k(x,t).
\end{equation}
Of course, the zero-order terms in Eqs.~(\ref{ansatz}) and (\ref{ansatz2})
 do not generate a torque.
This expansion naturally suggests to define the adiabatic spin torque
as ${\bf T}_1$, the dominant piece for sufficiently
slow and smooth variation of ${\bf n}(x,t)$.  
As remarked above, this definition differs from standard usage 
but provides a systematic way to classify different spin torque 
contributions.

\subsection{Adiabatic spin torque}

Let us now compute the first-order spin torque ${\bf T}_1$.
The derivatives of ${\bf s}_1$ or ${\bf J}_1$ are of second
order and can be dropped here.
Inserting the Ansatz (\ref{ansatz}) and (\ref{ansatz2}) 
into Eqs.~(\ref{eqmsja}) and (\ref{eqmsjb}) then gives 
\begin{eqnarray} \label{firstordereq}
(\tau_{s}^{-1}- \Delta {\bf n} \times ) {\bf s}_1 
 & = & \chi_s \Delta \partial_t {\bf n} - J_s \partial_x {\bf n} 
 \equiv {\bf a}_1 ,
\nonumber \\
(\tau_{J}^{-1}- \Delta {\bf n} \times ) {\bf J}_1 
 & = & -J_s  \partial_t {\bf n} + v^2 \chi_s \Delta \partial_x {\bf n}
  \equiv {\bf b}_1 .
\end{eqnarray}
Here ${\bf J}_1$ and ${\bf s}_1$ completely decouple.  Exploiting
that the equation
$\lambda {\bf x} + {\bf y}\times {\bf x} = {\bf v}$
with real $\lambda$ and real $3$-vectors ${\bf x}$ and ${\bf y}$
is solved by 
\begin{equation}\label{auxrel}
{\bf x}= \frac{\lambda^2{\bf v}+ \lambda{\bf v}\times {\bf y}
+ ({\bf v}\cdot{\bf y}) {\bf y}}{\lambda(\lambda^2+|{\bf y}|^2)},
\end{equation}
it is straightforward to obtain 
\begin{eqnarray}\label{s1}
\Delta{\bf s}_1 &=& -{\bf a}_{1} \times {\bf n} + \beta {\bf a}_1 , \\
\label{j1}
\Delta {\bf J}_1 &=& -{\bf b}_{1} \times {\bf n} +\tilde \beta {\bf b}_1 ,
\end{eqnarray}
where we have introduced the dimensionless parameters 
\begin{equation}\label{beta}
\beta= \hbar/(\Delta\tau_{s}), \quad 
\tilde \beta= \hbar/(\Delta\tau_{J}). 
\end{equation}
In accordance with the weak-coupling Bloch-Redfield approach,
in numerical prefactors we have neglected terms of order 
$\beta^2$ and $\tilde \beta^2$
against unity. The parameters $\beta$ and
$\tilde\beta$ are small expansion parameters in this approach.

For the discussion of the adiabatic spin torque ${\bf T}_1$,
we only need ${\bf s}_1$ as given in Eq.~(\ref{s1}).
Inserting this into Eq.~(\ref{torq}) gives
the complete first-order spin torque,
\begin{equation} \label{torq1}
{\bf T}_1 = -J_s \partial_x {\bf n} + \beta J_s {\bf n}
 \times \partial_x {\bf n} +\Delta \chi_s \left[ 
 \partial_t {\bf n} -   
\alpha' {\bf n} \times \partial_t {\bf n} \right], 
\end{equation}
where $\alpha'= \beta$. 
The first term is the adiabatic spin torque (\ref{ad1}), and the 
second term is the `non-adiabatic' correction (\ref{liz}).
Moreover, the third term can be absorbed by renormalizing the 
gyromagnetic ratio $\gamma_0$ and related
parameters, cf.\  Ref.~\cite{li}. Finally, the last term gives an
additional contribution to the Gilbert damping that is
absorbed by a redefinition of $\alpha$. 
Our assumption of equal longitudinal and
transverse relaxation times,  $\tau_{x,y}=\tau_z =\tau_s$, 
cf. Appendix, strictly implies  that $\alpha'= \beta$. 
Under more
general conditions, however, this relation will not hold anymore 
\cite{tatara2,duine}. The approach presented here can
straightforwardly be generalized in this direction. 
Equation (\ref{torq1}) thus confirms and reproduces the results
of Zhang and Li \cite{li}, and effectively yields the first-order (adiabatic)
 spin torque in the form
\begin{equation} \label{firstord}
{\bf T}_1 = {\bf T}_{ad}+ {\bf T}_{ZL}.
\end{equation}

\subsection{Non-adiabatic spin torque}

Next we calculate the second-order terms in Eq.~(\ref{tanz}).  The second-order
contributions ${\bf s}_2$ and ${\bf J}_2$  are obtained by inserting the 
{\sl Ansatz} (\ref{ansatz}) and (\ref{ansatz2})  
into Eqs.~(\ref{eqmsja}) and (\ref{eqmsjb}), thereby exploiting 
Eq.~(\ref{firstordereq}). Dropping terms of higher than second order, the 
resulting equations can be solved again, with the result
\begin{eqnarray}\label{s2}
\Delta{\bf s}_2 &=& {\bf a}_{2} \times {\bf n} - \beta {\bf a}_2 , \\
\label{j2} 
\Delta {\bf J}_2 &=& {\bf b}_{2} \times {\bf n} -\tilde \beta {\bf b}_2 ,
\end{eqnarray}
with the auxiliary vectors 
\begin{eqnarray}
{\bf a}_2 & = & 
 \partial_t {\bf s}_1 + \partial_x {\bf J}_1 , \label{a2} \\
{\bf b}_2 & = & 
 \partial_t {\bf J}_1 + \partial_x {\bf s}_1 \, , \label{b2} 
 \end{eqnarray}
where again terms of order $\beta^2$ were discarded against unity 
in prefactors. 
We have also omitted terms $\propto {\bf n}$ resulting from Eq.\
(\ref{auxrel}) since these do not generate a spin torque. 
In fact, only ${\bf s}_2$ in Eq.~(\ref{s2}) is required. 
After straightforward algebra, we get
\begin{eqnarray} \label{force}
{\bf T}_2 &=& { \bf n}\times \Bigl( c_{tx} \partial_t \partial_x {\bf n} +
c_{tt} \partial_t^2 {\bf n} + c_{xx} \partial_x^2 {\bf n} \\ \nonumber
&+& {\bf n} \times [ d_{tx} \partial_t \partial_x {\bf n} +
d_{tt} \partial_t^2 {\bf n} + d_{xx} \partial_x^2 {\bf n} ] \Bigr)
\end{eqnarray} 
with the coefficients 
\begin{eqnarray} \label{coeffs}
 c_{tx}& = &  2 J_s/\Delta, \quad
c_{tt} =  c_{xx} / v^2 = -\chi_s,   \\  \nonumber
d_{tx} & = & -(3\beta + \tilde{\beta})J_s /\Delta, \\   \nonumber
d_{tt} & = &   2\beta \chi_s,  \quad d_{xx} =  v^2 \chi_s 
(\beta + \tilde{\beta})  . 
\end{eqnarray} 
We stress that as long as
the magnitude of the magnetization vector is conserved, ${\bf n}^2(x,t)=1$,
Eq.~(\ref{force}) represents the most general second-order spin torque
allowed by symmetry constraints.   
At this stage, let us compare to the results of Ref.~\cite{gerrit}, 
where a spin torque contribution ${\bf T}_{2}' = (J_s/\Delta) 
{\bf n} \times \partial_t \partial_x {\bf n}$ was reported.
This should match our $c_{tx}$-term, but differs by a factor 2. 
The reason for this difference is explained by noting that
Ref.~\cite{gerrit} employed the self-consistent Stoner description,
where one finds indeed a factor 2 difference to our $s$-$d$ model prediction
\cite{private}.  The other contributions to the second-order torque 
in Eq.~(\ref{force}), in particular the $d_{tx}$-term, 
have not been reported before.
Interestingly, the spin torque contains $J_s$-independent and linear-in-$J_s$
 terms only.  However, even the linear-in-$J_s$
spin torque (\ref{force}) leads to a nonlinear response
of the DW velocity as a function of $J_s$, see Sec.~\ref{sec3}.

The second order torque ${\bf T}_2$ in Eq.~(\ref{force})
introduces several new features into the LLG equation (\ref{llg}).  
First, the only terms proportional to the spin current $J_s$ are due
to $c_{tx}$ and $d_{tx}$. All other terms 
are less important when large spin currents are applied.
For instance, the $c_{xx}$-term can be combined with the
exchange term in the effective magnetic field
${\bf B}_{\rm eff}$ entering the LLG equation (\ref{llg}), see below in
 Eq.\ (\ref{beff}), i.e., it can be absorbed by a renormalization of 
the exchange constant $J$. 
Moreover, the $c_{tt}$ and $d_{tt}$-terms introduce
acceleration terms in the LLG equation. 
Since we are interested in current-driven spin torques here,
we will exclusively focus on the
$c_{tx}$ and $d_{tx}$ terms from now on.
We have checked explicitly that the results below do not qualitatively
change if the $d_{xx}, c_{tt}, d_{tt}$ terms are 
included in a perturbative scheme.
The second-order spin torque is thus taken as
\begin{equation}\label{secondtor}
{\bf T}_2 = \frac{2J_s}{\Delta} {\bf n} \times \left( \partial_t\partial_x
{\bf n} - \frac{3\beta+\tilde\beta}{2} {\bf n}\times \partial_t \partial_x
{\bf n} \right).
\end{equation}
This is a central result of our work.  Albeit the coefficient $d_{tx}$
is smaller by a factor $\approx 2\beta$ compared to $c_{tx}$,
it can qualitatively influence the magnetization dynamics, 
just as is the case for the first-order terms.  We believe
that Eq.~\eqref{secondtor}  should be included in micromagnetic simulations
of (for instance) DW dynamics.
Below we will see that these new terms lead, among other phenomena,
to a deformation of the shape of a DW,
as the spin torque now involves the time derivative of the
gradient of ${\bf n}$. 

\section{Domain wall motion}
\label{sec3}

In order to illustrate the effects of the new spin torque
terms (\ref{secondtor}), we now consider the problem of
current-induced domain wall motion.  Let us take a tail-to-tail DW, 
being a solution of the stationary LLG equation (\ref{llg}) with 
\begin{equation}\label{beff}
\gamma_0 {\bf B}_{\rm eff}[{\bf n}] =J \partial^2_x {\bf n} + K n_{z} 
\hat{e}_z - K_\perp n_{y} \hat{e}_y \, ,
\end{equation}
where $J$ is the ferromagnetic exchange coupling,  $K$ the 
longitudinal and $K_\perp$ the transverse 
anisotropy constant (incorporating the effects of 
demagnetizing fields). The resulting initial DW
configuration at time $t=0$ (when the spin current is 
switched on) with arbitrary center $x_0$ is then given by  
\begin{equation} \label{dw0}
{\bf n}(x,0)= 
\cosh^{-1} \left(\frac{x-x_0}{w}\right) \hat e_x 
+\tanh\left(\frac{x-x_0}{w}\right)\hat e_z,
\end{equation}
where $w=\sqrt{J/K}$ is the initial width of the wall.    
We then determine the time evolution of this configuration 
from the LLG equation in the presence
of a spin polarized current $J_s$ switched on at $t=0$, taking 
the full spin torque ${\bf T}={\bf T}_1+{\bf T}_2$,
see Eqs.~\eqref{firstord} and \eqref{secondtor}.

Due to the complexity of the resulting LLG equation, we 
have to solve it numerically.
To that end, we spatially discretize the LLG equation
into a discrete 1D spin chain, with spins ${\bf n}_i$ at $x=ia$ (integer $i$)
for lattice constant $a$. 
The resulting first-order ordinary differential equation (in time) 
is then solved numerically starting from the initial configuration (\ref{dw0}). 
To be specific, we choose a spatial grid with $x\in [0,1000a]$ and
scale time in units of $t_0=a/v$ (and energies correspondingly).
In these units, we take \cite{kramer} $J=3, K=0.1,
K_\perp=0.01,\Delta=0.2$, and $x_0=500a$. 
This implies the initial DW width $w\simeq 5.5 a$. 
Moreover, we choose $\alpha=\beta=\tilde{\beta}=0.03$. 
Let us briefly estimate the corresponding dimensionful parameters.
For this, we take the Fermi velocity of permalloy,
 $v=2.2 \times 10^5$ m/s \cite{py1}, with a unit
cell of volume $V_0=a^3=(1 \,\, {\rm nm})^3$. The time scale $t_0 =
4.5 \times 10^{-15}$~s follows.
The values for the magnetic constants are $J=4.4 \times 10^{-10}$~J/m,
$K=14.5 \times 10^{6}$~J/m$^3$, $K_\perp = 14.5 \times 10^5$~J/m$^3$,
and $\Delta= 0.18$~eV, resulting in the spin-flip time 
$\tau_s = 1/(\Delta \beta)=7.7 \times 10^{-13}$~s. Note that the spin
flip length for permalloy is $l_s \simeq 5$ nm \cite{py2} 
corresponding to a spin flip time of $\tau_s = 2.3 \times 10^{-14}$ s.  
A spin current density of $J_s=0.2$ then 
corresponds to a charge current density (with polarization $P=1$) of 
$I/A = 7 \times 10^{12}$~A/m$^2$. Typical current densities realized
in experiments are of this order of magnitude \cite{klaui,meier}. 
  
\begin{figure}
\includegraphics[height=65mm,keepaspectratio=true]{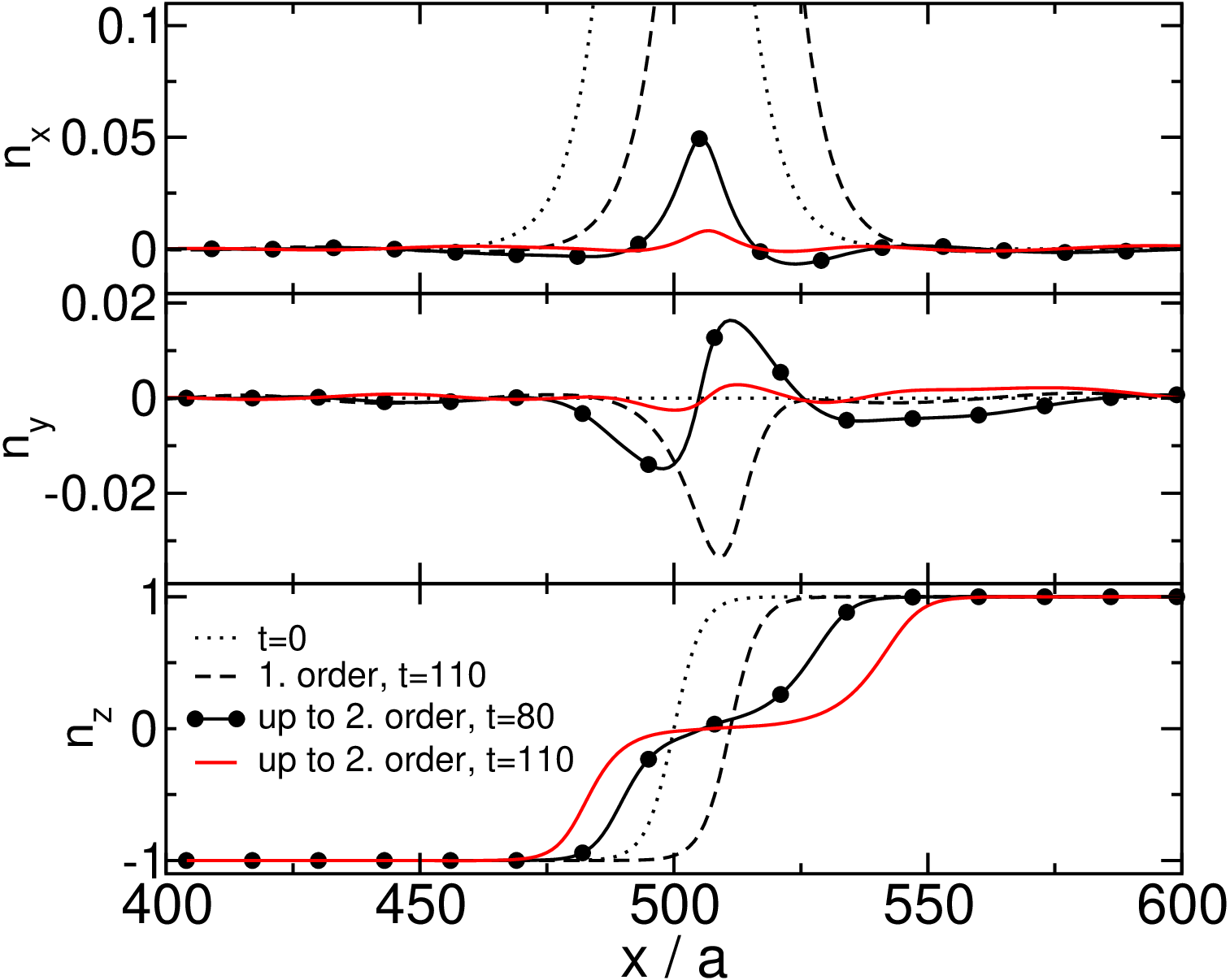}
\caption{(Color online) DW configuration at times $t=80 t_0$
[black solid line with filled circles] and $t=110 t_0$ [red solid line]
for $J_s=0.1$ starting from
the initial DW configuration (\ref{dw0}) [dotted line]. 
For comparison, the result of keeping only
${\bf T}_1$ is also shown (for 
$t=110 t_0$, dashed line).} \label{fig.1}
\end{figure}

The numerical results for the DW configuration at times 
$t=80 t_0$ and $t=110 t_0$ for $J_s=0.1$ 
are shown in Fig.\  \ref{fig.1}. 
As can be seen, the initial DW is moved to the right due to the presence
of the spin current. Importantly, the second-order spin torque 
leads to a considerable deformation of the initial shape of the DW,
where the magnetization rotates into the $x-y$ plane 
around the center of the DW.  This region grows with
time, but the DW center still moves along the direction of the current. 
For comparison, we also depict the results when only the first-order
spin-torque is taken into account.  This shows that the
smearing of the DW profile is caused by the second-order torque. 
The strong DW deformation is the reason why a simple Walker Ansatz
\cite{walker,lzad,li,tatara1} does not work.
In the Walker Ansatz,
one takes an adiabatic DW shape whose
center is moving with some velocity $V_{DW}(t)$, a spatially
independent out-of-plane angle $\phi(t)$, and a trial function for 
the polar angle $\theta ( [x- \int_0^t d\tau V_{DW}(\tau)] / w(t))$ with 
time-dependent DW width $w(t)$. Inserting such an Ansatz into the full
LLG equations yields four coupled differential equations
for three variables ($\phi(t),V_{DW}( t)$ and $w(t)$). 
Unfortunately, there is thus no consistent solution under a Walker Ansatz.

\begin{figure}
\includegraphics[height=55mm,keepaspectratio=true]{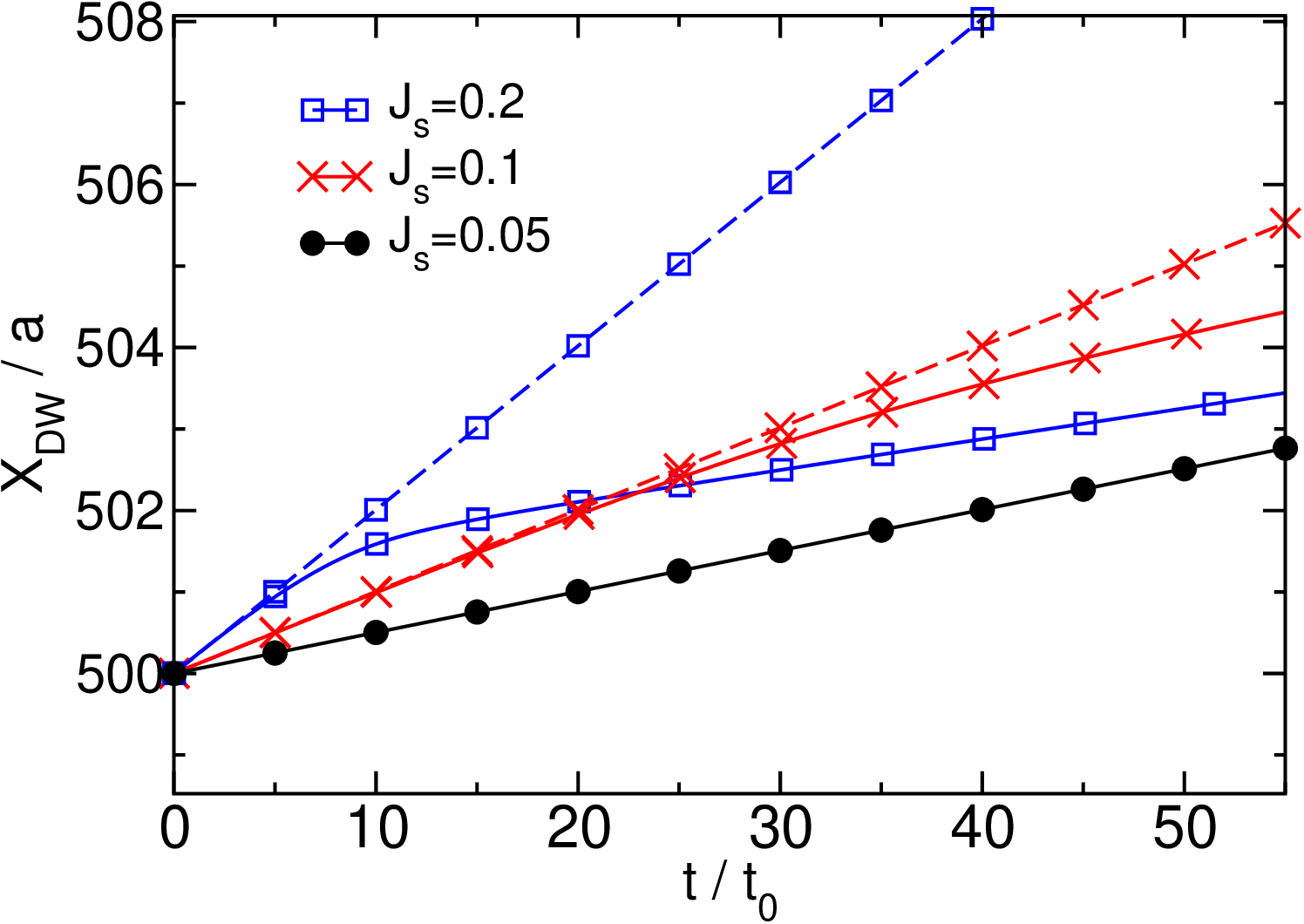}
\caption{(Color online) Time dependence of DW center for the
same parameters as in Fig.\  \ref{fig.1}, for 
 $J_s=0.05 (\bullet), 0.1 (\times)$ and $0.2 (\Box)$. Shown are the results
with the spin torque included up to first-order (dashed lines) and up to second
order (full lines). For $J_s=0.05$, both coincide in the shown regime of
time (black lines and filled circles).}
\label{fig.3}
\end{figure}

\begin{figure}
\includegraphics[height=55mm,keepaspectratio=true]{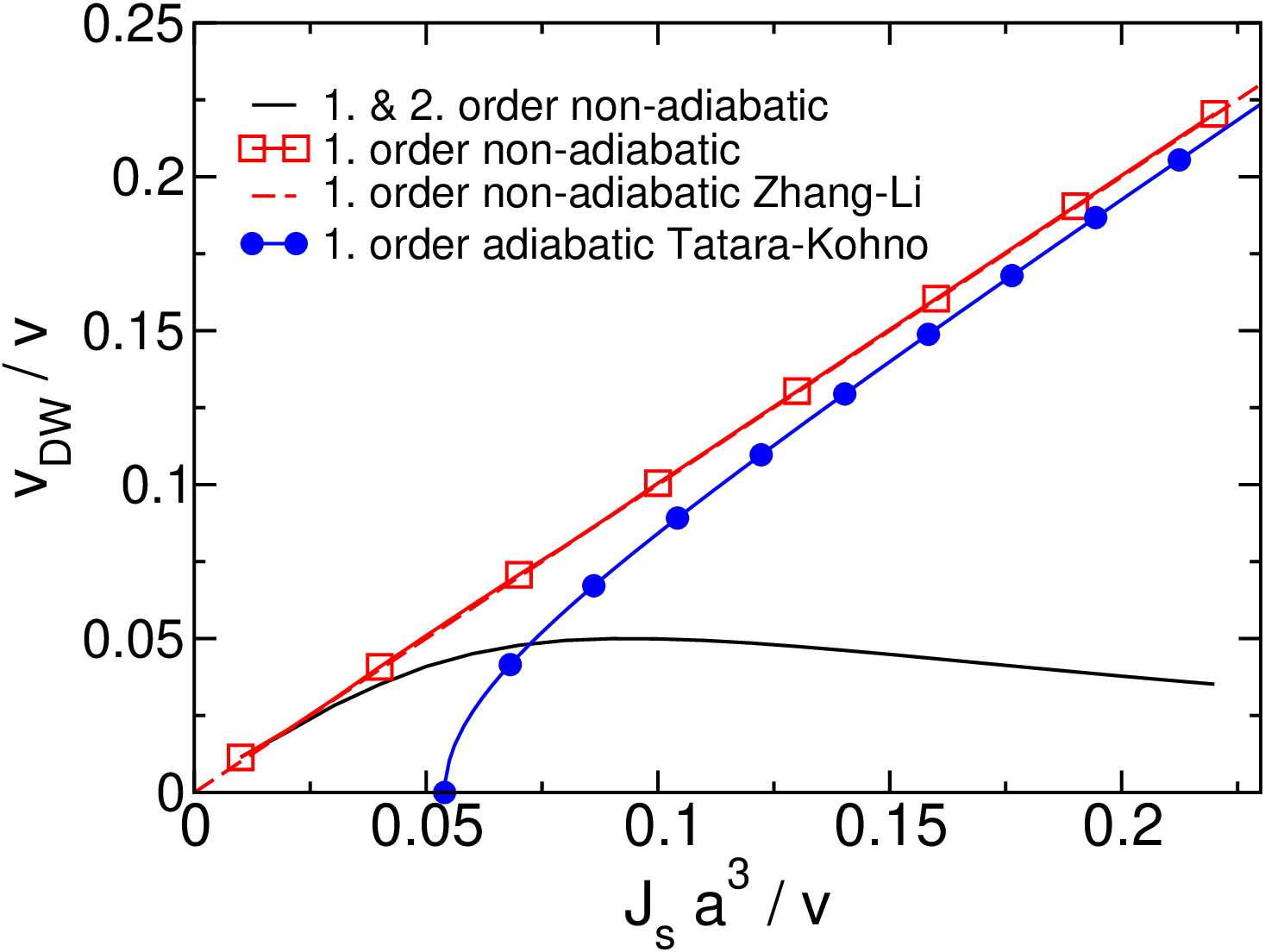}
\caption{(Color online) DW velocity vs.~spin
current density $J_s$ for same parameters as in Fig.~\ref{fig.1}. 
The black solid line gives the full numerical result, red empty
 squares the one when keeping only the first-order spin torque. The  
red dashed line gives the analytical result of Ref.\ \cite{li} 
(for the full first-order spin torque), the blue full line with filled
circles gives
the result of Ref.\ \cite{tatara1} where only the spin torque  (\ref{ad1}) 
is kept.  }
\label{fig.4}
\end{figure}

Despite the deformation of the DW, the
DW center $X_{DW}$ (defined as the zero of the $n_z$-component) displays a
uniform steady-state translational motion. This is shown in Fig.\
\ref{fig.3}, where  $X_{DW}(t)$ is shown for different $J_s$.
For the smallest value of the
spin current density, $J_s=0.05$, the first- and second-order results
for $X_{DW}(t)$ coincide in the shown regime of time. 
For larger spin currents, however, the DW center initially moves
mainly due to the first-order torque, and only after a transient time
the DW is deformed. The motion of the DW center is then
slower but still uniform, and a constant DW velocity can be extracted. 
At long times $V_{DW}(t) = d X_{DW}(t) / dt$ approaches
the constant value $V_{DW}$, shown in 
Fig.\  \ref{fig.4} as a function of $J_s$. 
We also show the DW velocity following from
only the first-order torque, where our numerical results reproduce
the known result $V_{DW}=\beta J_s / \alpha$ 
\cite{li}. As one can see, for small $J_s$, the
first- and second-order results merge. However, for
growing $J_s$, while the adiabatic (first-order) calculation gives a
steadily growing DW velocity with growing $J_s$, the second-order
(non-adiabatic) calculation indicates a non-monotonic dependence. 
This is a central result of our work, as it clearly shows that a
larger spin current does not necessarily increase the DW velocity.
Instead, it may lead to a DW deformation and a slower DW steady motion. 
This finding seems in qualitative agreement with experimental
observations where the measured DW velocities are typically smaller
than those predicted by the so far available first-order calculations
\cite{meier}. 
We note that deviations from the first-order result become noticeable
around $J_s \simeq 0.05$ which corresponds to a charge current density
of $I/A = 1.75 \times 10^{12}$ A/m$^2$ for our parameters (with
$P=1$). This value should be compared,  for instance, with the applied
current densities of $I/A = 1.0 \times 10^{12}$  A/m$^2$ in Ref.\
\onlinecite{meier} and of $I/A = 2.2 \times 10^{12}$  A/m$^2$ in Ref.\
\onlinecite{klaui}, indicating that the contribution of the
second-order spin torque can be significant already for intermediate
realistic current densities. 

In Fig.\ \ref{fig.4}, we also show the result of Ref.\
\cite{tatara1} where only the spin torque (\ref{ad1}) was included,
leading to $V_{DW}\simeq\sqrt{J_s^2 - J_{s,cr}^2}$ with the critical
threshold spin current  $J_{s,cr}= K_\perp w$.  The DW would then 
be pinned by the transverse anisotropy for small $J_s$. 
However, as has been discussed in Ref.\  \cite{li}, the full 
first-order spin torque lifts this pinning.

Finally, we note that our formalism allows to numerically 
solve the three coupled equations of motion for
 ${\bf n}(x,t), {\bf s}(x,t)$ and ${\bf
J}(x,t)$ directly. In particular, Eqs.\  
(\ref{llg}), (\ref{eqmsja}) and (\ref{eqmsjb}),
together with Eqs.\ (\ref{rels}) and (\ref{relj}), with the initial
configurations given in Eq.\ (\ref{dw0}) for ${\bf n}(x,0)$ and  
${\bf s}(x,0) = - \chi_s \Delta {\bf n}(x,0)$ and 
${\bf J}(x,0) = J_s {\bf n}(x,0)$,  constitute nine coupled first-order
ordinary differential equations which can be integrated by a simple
fourth-order Runge-Kutta scheme.  Since this scheme does not yield a simple
closed formula for the effective spin torque ${\bf T}$, however,
we refrain from this approach here.  

\section{Concluding remarks }
\label{sec4}

In this paper, we have studied the spin torque due to an applied
spin-polarized current entering the Landau-Lifshitz-Gilbert 
equation for the magnetization $-M_s {\bf n}(x,t)$ 
of a ferromagnetic metallic wire.
Spin relaxation of the itinerant
electrons has been included within a Bloch-Redfield formalism
adapted to this problem.  The equation of motion of the
itinerant electrons can be iteratively solved within a 
gradient expansion scheme for current-induced spin torque ${\bf T}$ entering
the LLG equation.  The first order in this expansion contains
only first-order derivative terms (in particular, $\partial_x {\bf n}$),
and recovers the known `adiabatic' result (\ref{ad1}) and
the `non-adiabatic' correction (\ref{liz}). 
We prefer to call both terms `adiabatic' here, since they are both of
first order in the gradient expansion. The correction term discovered
in Ref.~\cite{li} is in fact due to spin relaxation \cite{li,gerrit,tatara2}.
We have then explicitly derived all second-order contributions 
to the spin torque term.
As they involve spatial and time derivatives of the
magnetization, they generally induce a deformation of the magnetization
profile as compared to the
simple adiabatic shape. This has been illustrated for the example of
a tail-to-tail transverse domain wall. Despite the fact that 
the  DW gets distorted,
it still moves uniformly with constant terminal velocity. We
have shown that this velocity depends in a non-trivial way on the applied spin
current density $J_s$. For small $J_s$, it essentially coincides with the
known first-order result, which illustrates that the `adiabatic' 
pinning of the DW is
overcome by the $\beta$-term (\ref{liz}) in the spin torque. For larger spin
currents, the DW velocity does not grow further with growing $J_s$ but
instead diminishes again. 
Most likely, upon inclusion of even higher orders, this indicates a saturation
of the DW velocity for large $J_s$.
Importantly, a simple
 analytical expression for the leading non-adiabatic corrections
to the known current-induced spin torque ${\bf T}_{ad}+{\bf T}_{ZL}$ 
has been provided in our work, see Eq.~(\ref{secondtor}). We 
hope that this result will also find its significance in more elaborate
micro-magnetic simulations.  Finally, we remark that also 
the problem of pinning \cite{tatara1} of a DW by structural 
defects could  be included in  
our formalism. The generalization to incorporate thermal  effects
\cite{duine2,gerrit2} via a stochastic LLG equation appears possible 
as well.

\acknowledgments
We thank Gerrit Bauer, Rembert Duine, Mathias Kl\"aui, and Gen Tatara
 for useful discussions.
This work was supported by the SFB TR 12 of the DFG.

\appendix

\section{Relaxation kernels}
\label{app}

In this Appendix, we sketch the derivation of the 
relaxation kernels entering the equations of motion
for the spin density and spin current density of the itinerant
electrons.   
 We use the abbreviations ${\bf K}_s (x)\equiv {\bf s}(x)$ 
 and ${\bf K}_J (x)\equiv {\bf J}(x)$, and
start from the master equation for the reduced
density operator under the Markov approximation for our
time-dependent system Hamiltonian  $H_S(t)=H_0+H_{ex}(t)$.  
This master equation has been derived in
detail in Ref.~\cite{abragam}, and can be 
taken over directly.  
The derivation assumes that the system is linearly 
coupled to a collection of environmental harmonic 
oscillators (with `bath' Hamiltonian $H_B$ and bath temperature $T$)
via a bilinear system-bath coupling $H_{SB}$. 
Specifically, we take 
\begin{equation}\label{hsbapp}
H_{SB} = \sum_{q=x,y,z; \lambda=s,J} \int dx B_{\lambda,q}(x,t)
 K_{\lambda,q}(x) ,
\end{equation}
 with fluctuating magnetic fields $B_{\mu,q}(x,t)$ produced by the
harmonic bath.  A key quantity is the (free) bath correlation function
 $\overline{B_{\mu,q}(x,t)B_{\nu,q'}(x',t-\tau)}$, or its Fourier transform,
 the spectral density \cite{weiss}. 
We will assume the spectral density to be real; neglecting
 its imaginary part only introduces small Lamb frequency shifts 
\cite{louisell,slichter,abragam,may} of no interest here.  
Thus we define
\begin{eqnarray} \label{kcor}
 k_{qq',\mu \nu}(x,x',\omega) &=&\frac{1}{2}\int_{-\infty}^{\infty}d\tau 
\cos(\omega\tau)\\ \nonumber &\times&
 \overline{B_{\mu,q}(x,t)B_{\nu,q'}(x',t-\tau)},
\end{eqnarray}
which is independent of $t$.
For our purpose, it is sufficient to assume a spatially 
independent isotropic spectral density which
yields an $x$-independent correlator 
$k_{q,\mu \nu}(\omega)=k_{qq',\mu \nu}(x,x',\omega)\delta_{qq'}$. 
The coupling $H_{SB}$ is now assumed weak enough to justify a 
lowest-order perturbative scheme. Moreover, the bath autocorrelation 
time $\tau_c$ should be sufficiently short to justify the 
standard Markov (no memory) approximation. 
Our starting point is thus the Markovian master equation for the 
reduced density matrix $\rho(t)$ of the system  \cite{abragam},
\begin{widetext}
\begin{eqnarray}\label{finmme}
\frac{d\rho(t)}{dt} & = & -i [H_S(t),\rho(t)] 
-  \frac{1}{2}\int_{-\infty}^{\infty}
  \ d \tau \overline{[H_{SB}(t),[e^{-iH_{S}(t)\tau}H_{SB}(t-\tau)
e^{iH_{S}(t)\tau} ,\rho(t)-\rho_S(t)]]}  ,
\end{eqnarray}
\end{widetext}
where the time-dependent Boltzmann density operator is given by 
\begin{equation}\label{instboltz}
\rho_S(t) = \frac{e^{-H_S(t)/k_B T}}{\mathrm{tr}
e^{-H_S(t)/k_B T}} .
\end{equation}
The system thus relaxes to the instantaneous Boltzmann distribution
(\ref{instboltz}) which contains the full time-dependent magnetic
field generated by ${\bf n}(x,t)$ in $H_{ex}(t)$. 

${}$From this result, we can then derive the 
Bloch-Redfield equations, i.e., the equations of
motion for the quantum statistical 
expectation values $\overline{\langle {\bf s}(x,t) \rangle}$ and $\overline{
\langle {\bf J}(x,t) \rangle}$. 
(For notational simplicity, we omit the brackets and the overline 
$\overline{\langle \cdot \rangle}$
in most of the paper, but we always mean 
the quantum statistical expectation value. In this Appendix, 
we shall keep them for clarity.) 
Let us now  consider the $r$-th component of 
$\overline{\langle {\bf K}_\lambda(x,t) \rangle}$ ($\lambda=s, J$),
which obeys the equation of motion 
\begin{equation}\label{exval}
\frac{d}{dt} \overline{\langle K_{\lambda,r}(x,t) \rangle} 
= \sum_{\alpha \alpha'} \frac{d \rho_{\alpha \alpha'}(t) }{dt} 
\mel{\alpha'}{K_{\lambda,r}(x,t)}{\alpha} ,
\end{equation}
since there is no explicit time 
dependence present in the operator ${\bf K}_\lambda$.
We have employed a suitable complete set of eigenstates, 
 $H_S(t)\ket{\alpha(t)}=\alpha(t)\ket{\alpha(t)}$, where for simplicity
$\alpha$ denotes both the quantum numbers and the eigenenergy,
and $\sum_{\alpha}\ket{\alpha}\bra{\alpha}=1$ (we keep the
$t$-dependence of $\alpha$ implicit from now on).  
By inserting Eq.\ (\ref{finmme}) into Eq.\ (\ref{exval}), 
we obtain first a part simply yielding 
the coherent terms in 
Eqs.~(\ref{eqmsja}) and (\ref{eqmsjb}). 
Here we focus on the relaxation part and
discuss the term $\sim \rho(t)$ in Eq.\ (\ref{finmme}). The
other term involving  $\rho_S(t)$ follows from a simple
substitution.  Therefore, we have to calculate
\begin{widetext}
\begin{equation}\label{exval2}
 \frac{d}{dt} \overline{\langle K_{\lambda,r}(x,t) \rangle}_{rel} 
= -\frac12 \sum_{\alpha \alpha'} \int_{-\infty}^{\infty}
  \!\!d\tau \overline{\mel{\alpha}{[H_{SB}(t),
[e^{-iH_{S}(t)\tau}H_{SB}(t-\tau)e^{iH_{S}(t)\tau}
  ,\rho(t)]]}{\alpha'}} \mel{\alpha'}{K_{\lambda,r}(x)}{\alpha} .
 \end{equation}
\end{widetext}
Expanding the double commutator yields four terms, of which one
will be discussed in more detail. 
Inserting unities, we find such a term as
\begin{eqnarray*}
 &&  \sum_{\alpha \alpha' \beta \beta'} 
  \frac{1}{2}\int_{-\infty}^{\infty}
  \!\!d\tau \overline{\mel{\alpha}{H_{SB}(t)}{\beta} 
 \mel{\beta'}{H_{SB}(t-\tau)}{\alpha'}} 
  \\ &&
 \times e^{i(\alpha'-\beta')\tau}
  \mel{\beta}{\rho}{\beta'} \mel{\alpha'}{K_{\lambda,r}(x)}{\alpha} .
\end{eqnarray*}
Using the bath correlation functions introduced above,
this can be written as
\begin{eqnarray*}
 &&  \sum_{\alpha \beta q} 
  \sum_{\mu \nu} 
\int dx' dx'' \mel{\beta}{K_{\nu, q} (x'')}{\alpha}  \\ &&
 \times\mel{\alpha}{K_{\lambda,r}(x)K_{\mu, q} (x')\rho}{\beta}
  k_{q,\mu \nu}(\beta -\alpha)   .
 \end{eqnarray*}
Collecting all four terms then yields the relaxation part
\begin{eqnarray}
 && \lefteqn{
 \frac{d}{dt} \overline{\langle K_{\lambda,r}(x) \rangle}_{t,rel}  
=  \sum_{\alpha \beta q} \sum_{\mu \nu}
\int dx' dx''\mel{\beta}{K_{\nu, q} (x'')}{\alpha}}  \nonumber  \\
& & \times  \mel{\alpha}{[[K_{\lambda,r}(x),K_{\mu, q} (x')],\rho]}{\beta}
  k_{q,\mu \nu}(\beta -\alpha) ,
\label{exval5}
 \end{eqnarray}
where we exploit that $k_{q,\mu \nu}(\omega)$ is even in $\omega$.  

Applying Eq.~(\ref{exval5}) to the case $K_{s,r=z}\equiv s_z$ is 
straightforward upon using the Kac-Moody algebra
and cyclic invariance of the trace.
In the $q$-summation, the contribution for $q=x$ 
involves the matrix elements 
of the operator $s_x$.  States $\alpha$ and
$\beta$ joined by $s_x$ thus have $\beta-\alpha=\Delta$,
the Larmor frequency for this problem.
A similar reasoning applies to $q=y$,
 while $q=z$ does not contribute at all.
Collecting all pieces yields for the relaxation part 
\[
\frac{d}{dt}\overline{\langle s_z (x,t) \rangle}_{rel}
  = - \frac{1}{\tau_{s,z}} \overline{\langle s_z (x,t) \rangle} ,
\]
with 
\[
\frac{1}{\tau_{s,z}} = k_{x,ss}(\Delta)+ k_{y,ss}(\Delta) + 
v^2[k_{x,JJ}(\Delta)+ k_{y,JJ}(\Delta)] ,
\]
where we use $k_{q,\mu \nu}(\omega)=k_{q,\nu \mu}(\omega)$. For 
$r=x,y$, we proceed in the same way and find 
\[
\frac{d}{dt}\overline{\langle s_{x/y}(x,t) \rangle}_{rel} = -
 \frac{1}{\tau_{s,x/y}} 
\overline{\langle s_{x/y}(x,t) \rangle}
\]  
with 
\[
\frac{1}{\tau_{s,x/y}} = k_{y/x,ss}(\Delta) + k_{z,ss}(0)+  
v^2[k_{y/x,JJ}(\Delta) + k_{z,JJ}(0) ] .
\]
Assuming isotropy, $\tau_{s,x}\approx 
\tau_{s,y}\approx \tau_{s,z}\approx \tau_{s}$, and 
adding the part from the  stationary distribution \eqref{instboltz},
we finally obtain the relaxation term \eqref{rels} 
entering Eq.~(\ref{eqmsja}).
In the same manner, ${\bf \Gamma}_J$ follows from such a calculation.

\end{document}